\documentclass[aps,showpacs,floatfix]{revtex4}
\usepackage{graphics}
\usepackage{dcolumn}
\usepackage{amssymb}
\newcommand{\bea}{\begin{eqnarray}}
\newcommand{\eea}{\end{eqnarray}}
\newcommand{\be}{\begin{equation}}
\newcommand{\ee}{\end{equation}}

\def\be{\begin{eqnarray}}
\def\ee{\end{eqnarray}}
\def\bd{\begin{displaymath}}
\def\ed{\end{displaymath}}

\def\NP{Nucl. Phys. }
\def\PR{Phys. Rev. }
\def\PRL{Phys. Rev. Lett. }
\def\PL{Phys. Lett. }

\begin{document}
\title
{Cluster decay in very heavy nuclei in Relativistic Mean Field}
\author{Madhubrata Bhattacharya} 
\author{G. Gangopadhyay}
\email{ggphy@caluniv.ac.in}
\affiliation{Department of Physics, University of Calcutta\\
92, Acharya Prafulla Chandra Road, Kolkata-700 009, India}
\begin{abstract}
Exotic cluster decay of very heavy nuclei has been studied in the
microscopic Super-Asymmetric Fission Model. Relativistic Mean Field model 
with the force FSU Gold has been employed to obtain the densities of the 
cluster and the daughter nuclei. The microscopic nuclear interaction DDM3Y1,
which has an exponential density dependence,
and the Coulomb interaction have been used in the double folding model 
to obtain
the potential between the cluster and the daughter. Half life values have 
been calculated in the WKB approximation and the spectroscopic factors
have been extracted. The latter values
are seen to have a simple dependence of the mass of the cluster as has been 
observed earlier. Predictions
have been made for some possible decays.

\end{abstract}
\maketitle
                                                                                
\vskip 0.5cm

Decay of cluster of nucleons from very heavy nuclei was suggested by Sandulescu 
{\em et al.} \cite{Gr} in 1980. Subsequently, 
emissions of various clusters from very heavy nuclei leading to daughters 
around the magic nucleus $^{208}$Pb have been observed\cite{exp} experimentally. 
This new type of radioactivity is analogous to alpha decay where the decaying
particle tunnels through the potential barrier. 
A lot of work has been done by various authors to calculate the half life
for these exotic decay processes. Super-Asymmetric Fission Model (SAFM) has been used 
in \cite{safm} and Preformed Cluster Model in \cite{pcm} to describe the various
observed decays.

In the present work, we have used the microscopic SAFM. 
 In most of the previous works in this model, the potentials that have
been used were phenomenological in nature. Even in cases where the potential 
has been constructed from nuclear densities, the densities themselves were
obtained using phenomenological prescriptions. However,
microscopic densities obtained from mean field approaches may be expected to 
provide a better description of the densities and hence that of the process
of decay.  With this in mind, we have chosen Relativistic Mean Field (RMF) to 
calculate the densities and study cluster decay in the present brief report.

There has been a number of calculations for half life of cluster decay under 
the SAFM and we cite only a few recent ones.
Basu \cite{basu} has studied the nuclear cluster radioactivity 
also in the framework of a SAFM using a phenomenological density and the 
realistic M3Y interaction. Bhagwat and Gambhir \cite{Amiya} have used densities 
from RMF calculations using the force NL3\cite{NL3}. 
In the present work, we have used another form of the Lagrangian density 
for RMF to 
study the structure of the ground state of the nuclei. We  also have selected a 
form of the interaction with an
exponential density dependence, DDM3Y1, different from that used in Ref. 
\cite{Amiya}.  This particular form of
the interaction is consistent in the sense that it can also explain the 
low energy scattering in the double folding approach.

The interaction potential between the decaying cluster and the daughter 
nucleus has been obtained in the present work applying the double folding model 
by folding the densities of 
the cluster and the daughter nucleus using some suitable nuclear interaction
along with the Coulomb interaction. 
As already mentioned, usually the densities are obtained from phenomenological 
description while in the present work, we have employed 
RMF to obtain the densities of the cluster and the daughter nucleus.

RMF is now a standard tool in low energy nuclear structure, successfully 
reproducing various features such as ground state binding energy, deformation, 
radius, excited states, spin-orbit splitting, neutron halo, etc.
It is well known that in nuclei far away from the
stability valley, the single particle level structure undergoes certain
changes in which the spin-orbit splitting plays an important role.
Being based on the Dirac Lagrangian density, RMF is particularly suited to
investigate these nuclei because it naturally incorporates the
spin degrees of freedom. In the case of exotic decays, certain neutron rich 
clusters such as $^{26}$Ne, $^{30}$Mg and $^{34}$Si are emitted. 
We expect RMF to successfully describe the nucleon densities 
in these clusters.
There exist different variations of the Lagrangian density as
well as a number of different parameterizations in RMF. Recently, a new Lagrangian
density has been proposed\cite{prl} which involves
self-coupling of the vector-isoscalar meson as well as coupling between the
vector-isoscalar meson and the vector-isovector meson. The corresponding
parameter set is called FSU Gold\cite{prl}. This Lagrangian density has earlier 
been employed to obtain the proton nucleus interaction to successfully 
calculate the half life for proton radioactivity\cite{plb}. In this work also, 
we have employed this force.

In the conventional RMF+BCS approach for even-even nuclei, the Euler-Lagrange
equations are solved under the assumptions of classical meson
fields, time reversal symmetry, no-sea contribution, etc. Pairing is introduced
under the BCS approximation. Since the nuclear density as a function of 
distance is very
important in our calculation, we have solved the equations in co-ordinate space.The strength of the zero range pairing force is taken as 300 MeV-fm for both
protons and neutrons. 

The microscopic density dependent M3Y interaction (DDM3Y) may be obtained from a
finite range nucleon nucleon interaction by introducing a density dependent factor.
This class of interactions has been employed widely in the study of
nucleon-nucleus as well as nucleus-nucleus scattering, calculation of proton
radioactivity, etc.
In this work,
we have employed the interaction DDM3Y1 which has an exponential 
density dependence
\bea
v(r,\rho_1,\rho_2,E)=C(1+\alpha\exp{(-\beta(\rho_1+\rho_2)})
(1-0.002E)u^{M3Y}(r)\eea
used in Ref. \cite{Khoa} to study alpha-nucleus scattering. It uses the direct
M3Y potential $u^{M3Y}(r)$ based on the $G$-matrix elements of the
Reid\cite{Reid} NN potential and reproduces the saturation properties of cold 
nuclear matter.  The weak energy dependence was introduced\cite{Khoa1} to
reproduce the empirical energy dependence of the optical potential. 
The parameters used have the standard values {\em viz.} $C=0.2845$, $\alpha=3.6391$ 
and $\beta=2.9605$fm$^2$. Here $\rho_1$ and $\rho_2$ are the densities of 
the $\alpha$-particle and the daughter nucleus. and $E$ is the energy of the 
$\alpha$-particle per nucleon in MeV.
 This interaction has been folded
with the theoretical densities of cluster and the daughter nucleus
in their ground states using the code DFPOT\cite{dfpot}.
The assault frequency
has been calculated from the decay energy following Gambhir
{\em et al}\cite{gambhir}. The centrifugal barrier has been included in the 
cases where the $l$ value for the decay is known to be non-zero.

The decay probabilities have been calculated in the WKB approximation for
penetration of the barrier by the cluster.
No theoretical calculation can reproduce the $Q$-values for the decay very 
accurately. As the decay probability changes very rapidly with $Q$-value, we 
have taken the $Q$-values (and the decay energies) from experimental 
measurements.

The spectroscopic factors have been calculated as the ratios of the calculated 
half lives to the experimentally observed half lives. This represents the 
preformation factor and may be considered as the overlap of the actual ground
state configuration and the configuration representing the cluster coupled to
the ground state of the daughter.
Obviously it is expected to be much less than unity. The results for the spectroscopic factors are presented 
in TABLE \ref{tab1}.
Theoretical calculations have been performed for spectroscopic factors for some of these nuclei. Our results are comparable to those values. For example,
our calculated value for S of $^{212}$Po is $1.88\times10^{-2}$ as compared to 
theoretical
value $2.5\times10^{-2}$ deduced in \cite{th}.
A value of $3.1\times10^{-2}$ was obtained by Mohr\cite{Mohr} in a double folding model 
calculation using density from experimentally known charge distribution.

\begin{figure}
\resizebox{\columnwidth}{!}{\includegraphics{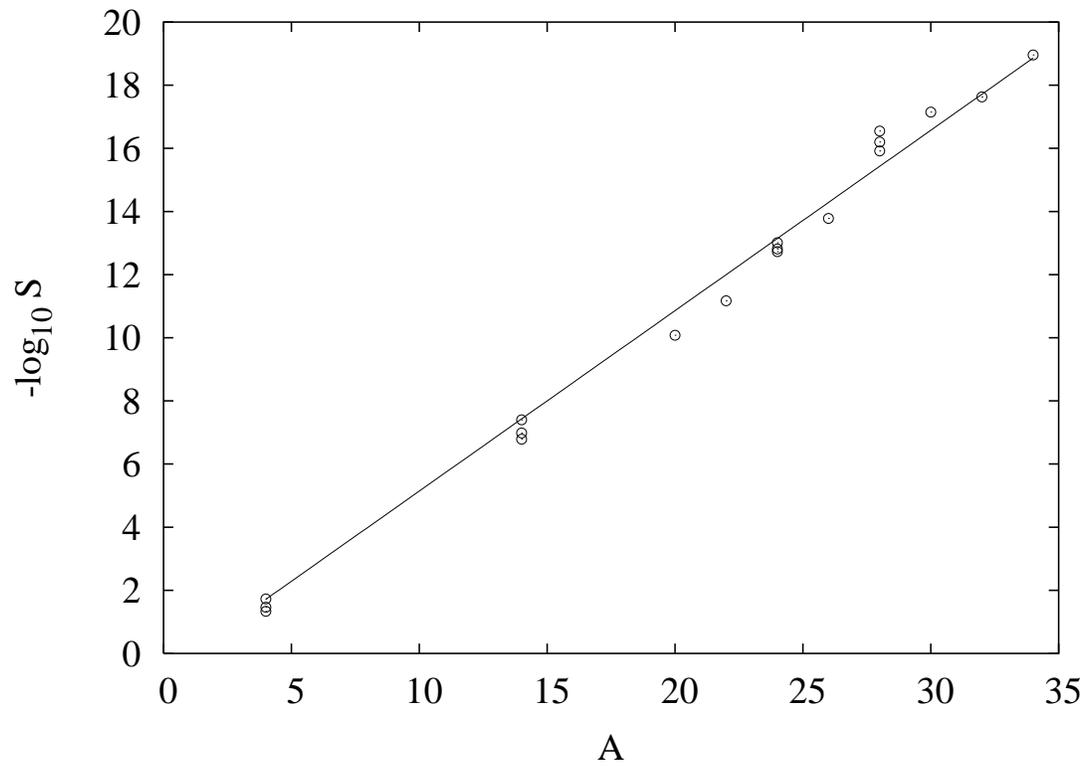}}
\caption{Negative of logarithm of spectroscopic factors ($S$) as a function of 
cluster mass number $A$ for even-even parents and daughters. 
\label{cl1}}
\end{figure}

It has been suggested\cite{cluster} that in the case of decay of heavy 
clusters, 
the spectroscopic factor may scale as 

\be S=(S_\alpha)^{(A-1)/3}\ee
where $A$ is the mass of the heavy cluster and $S_\alpha$ is the 
spectroscopic factor for the $\alpha$-decay. Thus a plot of  $\log_{10} S$ against 
$A$ should be a straight line. In Fig. \ref{cl1}, we have plotted the 
negative of $\log_{10} S$ for the decays where both the parent and the daughter are
even-even nuclei against the mass number of the cluster and plotted 
a best fit line. One can see that the points fall nearly on a straight 
line with the $S_\alpha$ value given by $1.93\times10^{-2}$. This is comparable to
the  value $1.61\times10^{-2}$ obtained by Poenaru {\em et al}\cite{Poenaru}.

\begin{figure}
\resizebox{\columnwidth}{!}{\includegraphics{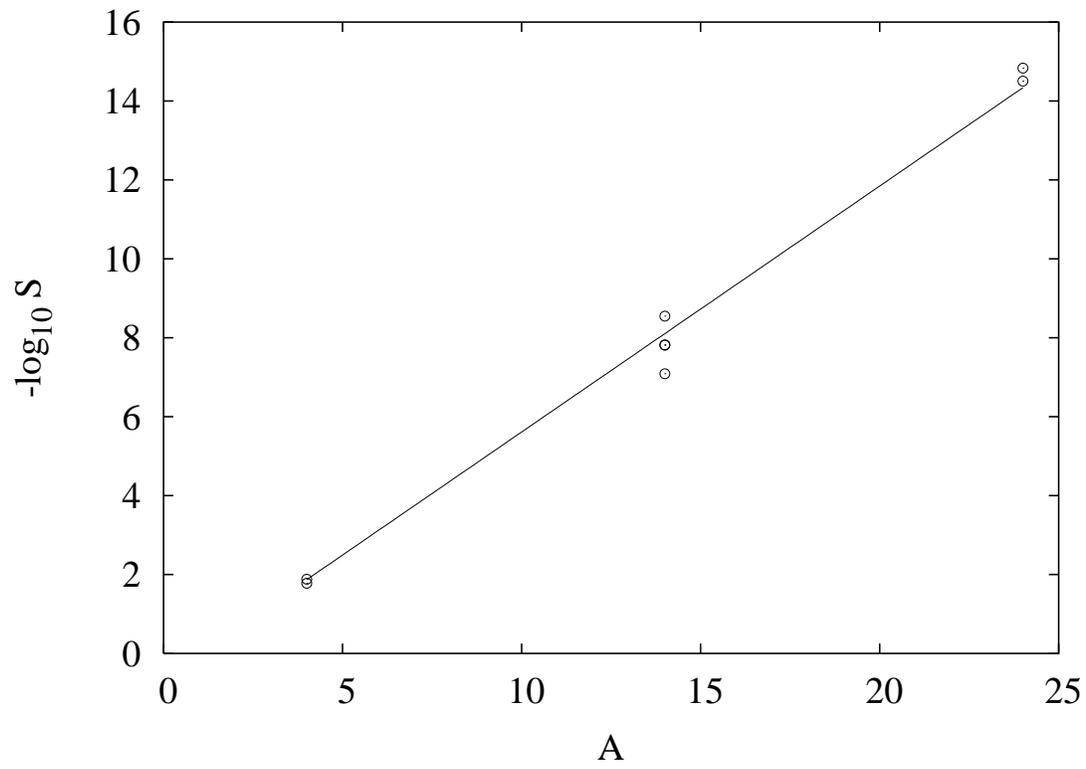}}
\caption{Negative of logarithm of spectroscopic factors ($S$) as a function of 
$A$ for odd mass parents and daughters. 
\label{cl2}}
\end{figure}

We have also extended our study to decays where both the parent and the daughter
nuclei have odd mass though the number of observed decays is rather small. The corresponding curve is shown in Fig. \ref{cl2}. 
Here the $S_\alpha$ value given by $1.35\times10^{-2}$.

 With such a good linear fit of the logarithm of spectroscopic factors with mass 
numbers, we have extended our scheme to calculate the half life of some other 
possible decays where
unambiguous  lifetime measurements are not yet available possible or where
there are possibilities of some decay taking place. Our results obtained with
the fitted values of $S$ from eqn. (1) are tabulated in TABLE \ref{tab2}
along with whatever experimental data are available. Except in the case of the
decay of $^{233}$U, the results are consistent with experimental observations.
Even in the case of $^{233}$U, the error is small.

To summarize, we have studied exotic cluster decay of very heavy nuclei in the 
microscopic SAFM. The densities of the cluster and the 
daughter nuclei have been obtained from RMF. The microscopic interaction DDM3Y1 
and the Coulomb interaction have been folded with these densities to obtain 
the potential between the cluster and the daughter. The lifetime have
been calculated in the WKB approximation and the spectroscopic factors
have been extracted.
The spectroscopic factors 
are seen to follow a simple rule as have been observed earlier. Predictions
have been made for some possible decays.

This work was carried out with financial assistance of the
Board of Research in Nuclear Sciences, Department of Atomic Energy (Sanction
No. 2005/37/7/BRNS).

\newpage
\begin{table}
\caption{Spectroscopic factor (S) of cluster decay obtained in the present 
calculation.\label{tab1}}
\center
\begin{tabular}{cccccc}\hline
Parent & Cluster & $Q$ (MeV) &log$_{10}$T$_{ex}$(s) &  S  \\\hline
 $^{212}$Po &  $^4$He   & 8.95~   & -6.52 & 1.88$\times10^{-2}$ \\
 $^{213}$Po &  $^4$He   & 7.833  & -5.44 & 1.67$\times10^{-2}$ \\
 $^{214}$Po &  $^4$He   & 7.833  & -3.78 & 3.45$\times10^{-2}$ \\
 $^{215}$At &  $^4$He   & 8.178  & -4.00 & 1.31$\times10^{-2}$ \\
 $^{221}$Fr & $^{14}$C & 31.317 & 14.52 & 1.50$\times10^{-8}$\\
 $^{221}$Ra &  $^{14}$C &  32.396& 13.39 & 1.55$\times10^{-8}$\\
 $^{222}$Ra &  $^{14}$C &  33.05~ & 11.00 & 1.64$\times10^{-7}$\\
 $^{223}$Ra &  $^{14}$C &  31.829& 15.20 & 2.85$\times10^{-9}$\\
 $^{224}$Ra &  $^{14}$C &  30.54~ & 15.92 & 1.04$\times10^{-7}$\\
 $^{225}$Ac &  $^{14}$C & 30.477& 17.34 & 8.14$\times10^{-8}$\\ 
 $^{226}$Ra &  $^{14}$C &  28.20~ & 21.34 & 3.97$\times10^{-8}$\\
 $^{228}$Th & $^{20}$O & 44.72~  & 20.72 & 8.37$\times10^{-11}$\\
 $^{230}$U & $^{22}$Ne & 61.40~  & 19.57 & 6.72$\times10^{-12}$\\
 $^{230}$Th & $^{24}$Ne &57.571 &  24.64 & 1.87$\times10^{-13}$\\
 $^{231}$Pa & $^{24}$Ne & 60.417 & 23.38 & 3.13$\times10^{-15}$\\
 $^{232}$U & $^{24}$Ne & 62.31~  & 20.40& 9.77$\times10^{-14}$\\
 $^{233}$U & $^{24}$Ne & 60.486 & 24.82 & 1.47$\times10^{-15}$\\
 $^{234}$U & $^{24}$Ne & 58.826 & 25.25 & 1.54$\times10^{-13}$\\
 $^{233}$U& $^{25}$Ne & 60.776 & 24.82 & 4.02$\times10^{-16}$\\
 $^{234}$U& $^{26}$Ne & 59.466 & 25.07 &  1.67$\times10^{-14}$\\
 $^{234}$U& $^{28}$Mg & 74.11~  & 25.74 &  6.30$\times10^{-17}$\\
 $^{236}$Pu& $^{28}$Mg & 79.67~  & 21.67&  2.83$\times10^{-17}$\\
 $^{238}$Pu& $^{28}$Mg & 75.912 & 25.70  &  1.21$\times10^{-16}$\\
 $^{238}$Pu& $^{32}$Si & 91.19~  & 25.28&  2.34$\times10^{-18}$\\
 $^{238}$Pu& $^{30}$Mg & 91.19~  & 25.28&  2.34$\times10^{-18}$\\
 $^{242}$Cm& $^{34}$Si &  96.509& 23.15 &  1.10$\times10^{-19}$\\
 \hline
\end{tabular}
\end{table}
\begin{table}
\caption{Half life values of cluster decay obtained in the present
calculation.\label{tab2}}
\center
\begin{tabular}{cccccc}\hline
Parent & Cluster & $Q$ (MeV) &log$_{10}$T$_{ex}$(s) & log$_{10}$T$_{th}$\\\hline
$^{224}$Th &  $^{14}$C & 32.929& &13.68\\
$^{226}$Th &  $^{14}$C & 30.596&&18.28\\
$^{224}$Th &  $^{16}$O & 46.481&&15.47\\
$^{226}$Th &  $^{18}$O & 45.727&$>16.8$&18.23\\
$^{232}$Th &  $^{24}$Ne & 54.497&&29.96\\
$^{236}$U & $^{24}$Ne & 55.945 &  &30.16 \\
$^{232}$Th &  $^{26}$Ne & 55.964&&28.57\\
$^{233}$U  & $^{28}$Mg & 74.226&$>27.6$&26.56\\
$^{237}$Np &  $^{30}$Mg & 74.817&$>27.6$&27.92\\
$^{240}$Pu & $^{34}$Si & 91.029&$>25.5$&26.48\\
$^{241}$Am &  $^{34}$Si & 93.926 &$>25.3$&26.25\\
\hline
\end{tabular}
\end{table}

\pagebreak

\pagebreak
\end{document}